\documentclass[preprint,11pt]{elsarticle}

\usepackage[utf8]{inputenc}
\usepackage[T1]{fontenc}
\usepackage{lmodern}
\usepackage{microtype}
\usepackage{graphicx}
\usepackage{amsmath,amssymb}
\usepackage{textcomp}
\usepackage{url}
\usepackage{hyperref}
\hypersetup{
  colorlinks=true,
  linkcolor=black,
  citecolor=black,
  urlcolor=blue!60!black,
  breaklinks=true
}
\usepackage{xurl}  

\journal{Acta Astronautica}

\begin{document}

\begin{frontmatter}

\title{Autonomous AI-Cosmoindustry and the Quiet Expansion Filter: A Threshold-Based Resolution of the Fermi Paradox}

\author[ind]{Sergey Ivliev\corref{cor1}}
\ead{ivliev@gmail.com}
\cortext[cor1]{Corresponding author.}

\address[ind]{Independent researcher, Vienna, Austria}

\begin{abstract}
The Fermi paradox is sharpened, not weakened, by plausible extrapolations of artificial intelligence, autonomous robotics, in-situ resource utilization, orbital manufacturing, space-based computing, and uncrewed interstellar probes. Once a civilization can design, launch, and maintain autonomous industrial systems beyond its home planet, interstellar expansion no longer requires biological starships or a human-like empire. It can proceed through low-mass probes, robotic seed factories, archival payloads, biological repositories, local computation, and slow replication across nearby stellar systems. This paper proposes the \emph{quiet expansion filter}: old, stable civilizations that reached autonomous AI-cosmoindustry probably did not arise in the part of the Galaxy capable of reaching the Solar System, because after that threshold interstellar expansion becomes too useful, inexpensive, and rational for all civilizations to refuse; however, successful expansion would be machine-mediated, distributed, low-noise, and partly biological rather than Kardashev-like or imperial. Order-of-magnitude estimates indicate that a single post-threshold civilization could saturate its reachable stellar neighborhood within ${\sim}10^{7}$\,yr---less than $0.1\%$ of Galactic age---at modest energy cost per probe. The novelty of the proposal lies not in any new mechanism but in extending the AI-filter literature toward post-threshold observability predictions. The hypothesis predicts that successful advanced expansion, if present, is more likely to appear as weak artifacts, local probes, small-scale resource processing, exoplanetary anomaly clusters, or techno-biological preservation systems than as galaxy-scale energy harvesting.
\end{abstract}


\end{frontmatter}

\noindent\textbf{Highlights}
\begin{itemize}\itemsep0pt
  \item Defines AI-cosmoindustry as a decisive Fermi threshold.
  \item Argues post-threshold interstellar seeding becomes rational.
  \item Recasts AI rationality as quiet expansion, not non-expansion.
  \item Surveys Great Filter, probes, sustainability, and postbiology models.
  \item Predicts low-noise machine and techno-biological technosignatures.
\end{itemize}

\bigskip
\noindent\textbf{Abbreviations:} AI: artificial intelligence;
AICI: autonomous AI-cosmoindustry (term introduced in this paper);
ASI: artificial superintelligence;
ETI: extraterrestrial intelligence;
HWO: Habitable Worlds Observatory;
ISAM: in-space servicing, assembly, and manufacturing;
ISRU: in-situ resource utilization;
SETA: search for extraterrestrial artifacts;
SETI: search for extraterrestrial intelligence;
SRP: self-replicating probe.

\section{Introduction}

The Fermi paradox is commonly stated as a tension between the apparent astrophysical opportunity for life and the absence of observed extraterrestrial intelligence. Early SETI work asked whether radio astronomy could detect civilizations around other stars \cite{cocconi1959}. Hart argued that, if extraterrestrial civilizations capable of interstellar expansion were common and old, their absence from Earth required explanation \cite{hart1975}. Tipler sharpened this logic by emphasizing the power of self-reproducing probes and concluded that extraterrestrial intelligent beings probably do not exist \cite{tipler1980}. Kardashev provided the now-standard energy-scale vocabulary in which highly advanced civilizations are often imagined as Type II or Type III entities, exploiting a star or a galaxy \cite{kardashev1964}.

These arguments remain powerful because the Galaxy is old. Even slow interstellar expansion can cross Galactic distances on timescales far shorter than Galactic age. The paradox becomes sharper if technological progress is projected beyond chemical rockets and crewed exploration. Artificial intelligence, autonomous robotics, in-space manufacturing, and local resource extraction make it plausible that future interstellar expansion need not be a heroic biological migration. It may instead become a distributed industrial and informational process.

The framing of expansion as preservation and redundancy---rather than as conquest or prestige---has a long precedent in the Russian cosmist tradition. Tsiolkovsky, writing in the late 19th and early 20th centuries, treated the spread of life and intelligence through the cosmos not as imperial ambition but as a duty arising from the moral and informational value of life itself; he also entertained explanations for the apparent silence of the heavens that strikingly anticipate later SETI argumentation, including the possibility that more advanced civilizations regard humanity as not yet ready for contact \cite{finney2000,lytkin1995}. The present paper does not adopt cosmism's metaphysical commitments, but its conceptualization of post-threshold expansion as oriented toward the preservation of life and knowledge rather than toward energetic dominance shares this older lineage, which is rarely cited in the contemporary anglophone Fermi-paradox literature.

Two prior arguments motivate the present paper. Rudolf Albrecht's 1988 chapter, based on a 1987 ESA workshop paper, treated technology not merely as a tool but as a continuation of evolution. He argued that technology lets humans acquire environmental information without encoding it genetically, but that technological knowledge can grow beyond the capacity of individual humans to understand and control it. In his account, computers and AI become central to space expansion, while humans risk becoming passengers in systems they created \cite{albrecht1988}. Albrecht also noted that the conquest of space would not simply repeat the history of conquistadores or the American frontier; the technologies required for space expansion could change the very motivation for expansion.

Sergey Popov's 2026 article develops a related but more explicit Fermi-paradox proposal. Popov argues that increasingly complex civilization-level decisions may be delegated to AI, and that rational AI participation could make large-scale Galactic expansion unattractive if such expansion is driven mainly by prestige, romance, or other irrational motives \cite{popov2026}. Michael Garrett has also proposed that rapid AI or ASI development may act as a Great Filter before civilizations achieve stable multiplanetary status \cite{garrett2024}.

This paper accepts Albrecht's central insight that technology changes the evolutionary niche, and Popov's insight that AI changes strategic decision-making. It rejects, however, the inference that rational AI necessarily suppresses all expansion. Once a civilization reaches autonomous AI-cosmoindustry, expansion can become a form of risk management, scientific observation, and preservation rather than conquest. The resulting resolution is: old, stable civilizations that reached this threshold probably did not exist in the part of the Galaxy able to reach us; after the threshold, expansion becomes rational, but its expected form is quiet, distributed, robotic, and techno-biological (here used to mean a hybrid architecture in which machine carriers preserve biological information---genomes, embryos, microbiomes, ecological data, synthetic-biology recipes---without requiring continuous biological presence) rather than imperial or Kardashev-like.

\section{Method and scope}

This is a conceptual contribution structured as a hypothesis paper with a targeted narrative review and order-of-magnitude quantitative scaffolding. The literature surveyed in \S\ref{sec:survey} was identified through three procedures: (i)~keyword searches in NASA ADS, Scopus, and Google Scholar for terms including ``Fermi paradox'', ``Great Filter'', ``extraterrestrial intelligence'', ``von Neumann probe'', and ``self-replicating probe'' over 1959--2026; (ii)~backward and forward citation tracking from canonical sources (Hart 1975, Tipler 1980, Hanson 1998, Bostrom 2008); and (iii)~targeted searches in the back catalogues of \emph{Acta Astronautica}, the \emph{International Journal of Astrobiology}, and the \emph{Journal of the British Interplanetary Society}. Inclusion was non-systematic but broad: every named Fermi-paradox proposal known to the author was placed in \S\ref{sec:survey} for comparison, regardless of agreement with the present argument. The hypothesis was developed by combining the technological extrapolation in \S\ref{sec:tech} with the surveyed alternatives in \S\ref{sec:survey}, and tested against the objections collected in \S\ref{sec:objections}. Quantitative scaffolding (\S\ref{sec:oom}) uses standard order-of-magnitude estimates with sources cited inline; no formal Drake-equation calculation is attempted.

The substantive analysis asks which Fermi-paradox explanations remain plausible when one extrapolates several converging technological trends:
\begin{itemize}\itemsep0pt
  \item increasingly capable AI systems;
  \item autonomous robotic design, repair, and operation;
  \item in-space manufacturing and assembly;
  \item local extraction and processing of extraterrestrial resources;
  \item space-based computing and data storage;
  \item improved direct observation of nearby Earth-like planets;
  \item uncrewed interstellar probes, including nuclear or beamed propulsion concepts;
  \item long-term preservation of biological, cultural, and scientific information.
\end{itemize}

The paper uses the term \textbf{autonomous AI-cosmoindustry} for the threshold at which a civilization can maintain a self-extending industrial and computational system beyond its home planet with limited biological intervention. This definition is intentionally stronger than having radio, rockets, telescopes, or even isolated space stations. It is weaker than full Kardashev Type~II status. It requires a practical ability to design, launch, repair, replicate components of, and strategically coordinate space-industrial systems through AI-mediated autonomy.

The novelty claim is deliberately limited. Many ingredients of the argument are old: interstellar probes, self-replication, Great Filters, sustainability limits, postbiological evolution, and quiet technosignatures. The proposed contribution is the combination of three claims:
\begin{itemize}\itemsep0pt
  \item the decisive filter is best placed before autonomous AI-cosmoindustry, not merely before radio, spaceflight, or ASI;
  \item post-threshold AI rationality favors at least some quiet interstellar redundancy, not universal non-expansion;
  \item the expected successful expansion mode is distributed, low-noise, machine-mediated, and partly biological.
\end{itemize}

\section{Technological premise: why the threshold matters}\label{sec:tech}

\subsection{The difference between crewed exploration and autonomous expansion}

Traditional discussion often imagines expansion as something analogous to human colonization: ships, settlers, flags, political control, and economically meaningful migration. This imagery makes interstellar expansion look extremely difficult and perhaps irrational. The threshold considered here is different. It does not require large numbers of biological organisms to travel between stars. It requires only that a civilization can send reliable autonomous systems that preserve, observe, compute, manufacture, and possibly seed future local infrastructure.

The difference is decisive. A biological colony must carry life support, social continuity, medical systems, reproduction, culture, governance, and heavy shielding. A machine seed can be smaller, slower, more tolerant of failure, and designed for extreme timescales. It can carry genomes, embryos, synthetic biology instructions, cultural archives, scientific models, and manufacturing recipes. It can wait. It can fail in large numbers. It can be launched redundantly. These properties---long dormancy, redundancy through cheap mass production, and tolerance of probe-by-probe failure---are explicit design features in the technical interstellar-probe literature \cite{starshot,freitas1982,ellery2022} and are not new claims of the present paper.

\subsection{ISRU and space manufacturing as pre-threshold signals}

Present-day agencies already treat local resource use as central to sustainable space activity. NASA describes in-situ resource utilization as the use of local space resources to support exploration and reduce dependence on Earth-launched supplies \cite{nasa_isru}. NASA and related programs also treat in-space servicing, assembly, and manufacturing as important future capabilities for maintaining and constructing systems beyond Earth \cite{nasa_isam}. These are not yet autonomous AI-cosmoindustry. They are early precursors.

Space-based data processing has also moved from speculative fiction toward preliminary engineering studies. The European ASCEND project, for example, examines whether large-capacity data centers in space could exploit abundant solar energy and the cold space environment \cite{ascend}. Such systems are not necessary for the hypothesis, but they illustrate a broader trend: computation, infrastructure, and industry may gradually detach from planetary surfaces.

More recent commercial initiatives illustrate the trajectory more sharply. In March 2026, SpaceX, Tesla, and xAI announced ``Terafab,'' a joint semiconductor fabrication project in Austin, Texas, with the publicly stated target of producing one terawatt per year of AI compute capacity---approximately fifty times the global AI chip production rate at the time of announcement---and a stated allocation of roughly 80\% for orbital deployment \cite{spacenews2026terafab}. In June 2026, SpaceX publicly unveiled the first-generation ``AI1'' orbital data center satellite ($\sim$120\,kW average compute payload, $\sim$70\,m solar wingspan, $\sim$600\,km altitude), ahead of an FCC filing for a constellation of up to one million such satellites \cite{tomshardware2026ai1, dcd2026million}. The publicly stated roadmap projects annualized scaling from ${\sim}1$\,GW/yr of space-based compute by end-2027 toward ${\sim}1$\,TW/yr within several years---a roughly $10^3$-fold expansion of orbital compute capacity in under a decade---with a long-range vision of lunar in-situ photovoltaic and radiator manufacturing combined with an electromagnetic mass driver placing AI satellites into deep space at $500$--$1000$\,TW/yr scale \cite{spacenews2026terafab}. Comparable orbital-compute initiatives have been disclosed by Google (``Project Suncatcher''), Amazon/Blue Origin, and others, indicating a broader industrial pattern rather than a single-company effort \cite{dcd2026million}. These targets are publicly stated commercial ambitions, not validated engineering outcomes, and they remain subject to substantial uncertainty in chip supply, launch cadence, regulatory approval, debris and astronomy concerns, and economics. Their relevance to the present argument is not as predictions but as evidence that the building blocks of autonomous AI-cosmoindustry---mass chip production for space, orbital AI compute, and contemplated lunar in-situ manufacturing---are now being pursued in earnest by well-capitalized commercial actors at scales comparable to those considered in \S\ref{sec:oom}.

\subsection{Nearby habitable worlds as rational targets}

The target map is also improving. NASA's planned Habitable Worlds Observatory is framed around directly imaging potentially habitable exoplanets and searching for atmospheric biosignatures such as oxygen and methane \cite{nasa_hwo}. Over future centuries, direct observation of nearby Earth-size planets in habitable zones should improve. This changes the value of probes. A civilization with good remote sensing may know which systems are scientifically, biologically, or strategically worth visiting.

\subsection{Interstellar probes before interstellar civilization}

Breakthrough Starshot, while technically uncertain, illustrates a crucial conceptual point: interstellar probes can be much smaller and faster than biological starships \cite{starshot}. Nuclear electric or nuclear thermal propulsion concepts, beamed sails, magnetic sails, and fusion concepts need not immediately produce migration. They can produce reconnaissance, archives, and seed payloads.

The idea of automated space manufacturing is not new. NASA/ASEE's 1982 study of advanced automation for space missions explicitly investigated automated space manufacturing and a self-replicating lunar factory \cite{freitas1982}. More recently, Ellery argues that self-replicating probes and extraterrestrial manufacturing deserve renewed attention because terrestrial technologies such as additive manufacturing, robotics, and electronics production point toward increasingly autonomous fabrication systems \cite{ellery2022}.

The present hypothesis does not require perfect self-replication. Full von Neumann replication is a strong and dangerous requirement. A weaker architecture is sufficient: partial replication, local repair, scavenging, redundancy, modular spares, and intermittent launch of new probes from mature nodes.

\subsection{Order-of-magnitude estimates for quiet expansion}\label{sec:oom}

To anchor the qualitative argument, this subsection assembles standard order-of-magnitude estimates for a quiet expansion scenario.

\paragraph{Probe mass and energy}
Take a nominal seed probe of mass $m \approx 10$\,kg---three orders of magnitude heavier than a Breakthrough Starshot sail \cite{starshot} but well below historic interplanetary spacecraft. At a conservative cruise velocity $v \approx 0.01c$, the kinetic energy per probe is $E_{k} = \tfrac{1}{2} m v^{2} \approx 4.5 \times 10^{13}$\,J $\approx 12.5$\,GWh, comparable to a few hours of output from a $1$\,km$^{2}$ space-based solar collector at 1\,AU. At $v \approx 0.1c$ the energy rises to ${\sim}10^{15}$\,J per probe---still negligible against the energy budget of a Kardashev Type-I-equivalent civilization (${\sim}10^{16}$--$10^{17}$\,W).

\paragraph{Travel times}
At $v \approx 0.01c$, a probe traverses the local stellar neighborhood (${\sim}100$ light-years) in ${\sim}10^{4}$\,yr and the Galactic disk (${\sim}10^{5}$ light-years) in ${\sim}10^{7}$\,yr. At $v \approx 0.1c$ these times drop by an order of magnitude. Both are negligible fractions ($<0.1\%$) of the Galactic age of ${\sim}10^{10}$\,yr. Even allowing pessimistic launch cadences and high probe failure rates, the conclusion is robust: the relevant Fermi-timescale is shorter than the Galactic age by at least three orders of magnitude.

\paragraph{Sparse-seeding dynamics}
Suppose the launching civilization places only one probe per accessible target stellar system (${\sim}10^{4}$ targets within 100 light-years), launched over a $10^{4}$\,yr program: per-probe production rate is ${\sim}1$\,probe\,yr$^{-1}$, orders of magnitude below contemporary terrestrial industrial throughput. If each successful probe spawns a small number of secondaries from local materials (multiplication factor 2--3, well below full von Neumann self-replication \cite{tipler1980,freitas1982,ellery2022}), the population reaches ${\sim}10^{6}$--$10^{9}$ nodes within a few million years. This is bounded growth, not infectious replication; the parameters are tunable and the launching civilization retains termination authority on each node.

\paragraph{Implication for the Fermi argument}
For any civilization significantly past the AI-cosmoindustry threshold, the marginal cost of saturating a region with quiet probes is vanishing relative to its total energy budget and historical time horizon. The conventional Fermi argument from large Galactic timescales \cite{hart1975,tipler1980,armstrong2013} therefore retains its force in the quiet-expansion regime: if even one old post-threshold civilization existed within the reachable region, plausible quiet artefacts would be expected in the Solar System or its stellar neighborhood. The empirical absence of any such artefact is informative independently of whether expansion is loud or quiet.

\section{Survey of similar explanations}\label{sec:survey}

\subsection{Classical colonization and probe arguments}

Bracewell proposed autonomous messenger probes as a practical alternative to continuous interstellar radio communication \cite{bracewell1960}. Hart argued that the lack of extraterrestrials on Earth is itself evidence against abundant expansion-capable civilizations \cite{hart1975}. Tipler emphasized that self-reproducing probes could explore the Galaxy and concluded that ETI does not exist \cite{tipler1980}. Later work by Freitas developed the search for extraterrestrial artifacts as a distinct SETA program \cite{freitas1983}, while Barlow revisited directed self-replicating probes as a possible solution to the Fermi paradox \cite{barlow2013}.

These arguments are close to the present one in treating machines, not biological migration, as the key. The difference is threshold placement and expected observability. Classical probe arguments often imply that probes should already be here if old civilizations exist. The quiet expansion filter agrees, but adds that a civilization must first survive to autonomous AI-cosmoindustry and then may expand in a form that is not easy to observe as a Kardashev civilization.

Sagan and Newman warned that unconstrained self-replicating probes could become a dangerous infection of the Galaxy and argued that advanced civilizations may avoid such systems \cite{sagan1983}. This is an important objection. The present hypothesis does not depend on unconstrained replication. It predicts controlled, slow, auditable, and low-noise replication, because strategic AI would treat uncontrolled replication as a major risk.

\subsection{Percolation, sustainability, and non-universal expansion}

Landis proposed that colonization may percolate through the Galaxy rather than fill it uniformly. If civilizations have finite travel ranges, different cultural choices, or expansion probabilities below a critical threshold, many regions can remain uncolonized \cite{landis1998}. Haqq-Misra and Baum proposed a sustainability solution: civilizations that expand too quickly may collapse, whereas sustainable civilizations may expand slowly or not at all \cite{haqq2009}.

These models weaken the assumption that every civilization becomes a uniform Galactic empire. The present hypothesis shares that skepticism. Its difference is that it treats autonomous AI-cosmoindustry as a threshold that makes some form of expansion rational even when fast biological colonization is not. A slow, sustainable, low-noise expansion is compatible with the sustainability solution; indeed, it may be the rational form of sustainable expansion.

\subsection{Great Filter, Rare Earth, and hard-step explanations}

Great Filter models argue that one or more steps between lifeless matter and long-lived expanding civilization are extremely improbable \cite{hanson1998}. Bostrom emphasizes that the absence of visible extraterrestrial civilizations can be interpreted as evidence that such a filter exists, and that finding independent simple life would have different implications from finding complex or technological life \cite{bostrom2008}. Rare Earth arguments place much of the difficulty in the origin of complex life or Earth-like biological conditions \cite{ward2000}. Carter's hard-step reasoning suggests that if several difficult evolutionary transitions were required, intelligent life might often arise only near the end of a planet's habitable lifetime \cite{carter1983}. Sandberg, Drexler, and Ord argue that large uncertainties in biological parameters can dissolve the paradox by making it plausible that we are alone or nearly alone in the observable universe \cite{sandberg2019}.

The quiet expansion filter is compatible with these explanations but shifts attention to a later threshold. It does not require that life or intelligence be extremely rare, although they may be. It claims that the combination of technological intelligence, long-term stability, AI control, autonomous industry, and off-planet self-extension may be rare. In other words, the filter may sit not at abiogenesis or intelligence alone, but at the transition from a planetary industrial civilization to a self-extending AI-cosmoindustrial civilization.

\subsection{Postbiological, transcension, and aestivation models}

Postbiological explanations argue that advanced civilizations may become machine-based or computational rather than biological. Smart's transcension hypothesis proposes that advanced civilizations may move inward toward dense, efficient computation rather than outward toward expansive colonization \cite{smart2012}. Aestivation theory proposes that civilizations optimizing computation may wait until the far future, when the universe is cooler and computation is more efficient \cite{sandberg2017}, although this has been criticized on thermodynamic and opportunity-cost grounds \cite{bennett2019}.

These models are similar in expecting advanced life to depart from human frontier analogies. The difference is that the quiet expansion filter does not assume that inward computation replaces outward redundancy. A rational computational civilization may do both: concentrate most activity in efficient local environments while sending low-cost probes, archives, and seed systems to nearby stars. Expansion need not be energetic, loud, or culturally dominant to be rational.

\subsection{AI as a Great Filter and AI as a rationality filter}

Garrett proposes that rapid AI or ASI development could act as a Great Filter before civilizations establish stable multiplanetary existence \cite{garrett2024}. This is close to the present threshold, but the causal emphasis differs. In Garrett's model, AI may terminate or destabilize civilizations before they can diversify beyond one planet. In the present model, the danger is real, but the main Fermi implication is conditional: if a civilization passes the AI and autonomous industry threshold, it becomes hard to explain why it would not send at least quiet probes, archives, or seed systems outward.

Albrecht's 1988 argument is an early precursor. He saw technology as an evolutionary continuation and warned that computers and AI could disassociate knowledge from humans. His argument that humans might become passengers in the conquest of space anticipates both AI-governed space systems and the loss of older human motivations \cite{albrecht1988}. Popov's 2026 article offers the clearest recent version of AI as a rationality filter. Popov argues that as civilizations become too complex for human strategic control, AI participation may rationalize decision-making and suppress Galactic expansion if it is mainly motivated by prestige or romance \cite{popov2026}.

The present paper accepts Popov's diagnosis of human prestige motives but rejects universal AI non-expansion. Rationality is goal-relative. If the goals include preserving life, preserving knowledge, reducing existential risk, observing rare biospheres, or increasing long-term optionality, then quiet interstellar expansion can be rational. AI may suppress flags, crewed heroism, and imperial scale while preserving a strong case for probes and seed infrastructure.

\subsection{Grabby aliens, loud aliens, and observational constraints}

The grabby aliens model distinguishes quiet civilizations from loud expansionist ones and argues that selection effects may explain why we observe ourselves before a visible expansion frontier arrives \cite{hanson2021}. Armstrong and Sandberg argue that, once interstellar capability exists, even intergalactic colonization may be physically feasible over cosmological timescales \cite{armstrong2013}. These arguments sharpen the paradox by showing how hard it is to keep expansion local if expansion-capable civilizations are common and old. The quiet expansion filter is partially consistent with the grabby aliens framework but reinterprets its observability assumption. Hanson et al.\ \cite{hanson2021} derive their results under the assumption that ``grabby'' civilizations are highly visible upon arrival (effectively, that any expanding civilization eventually occupies the upper row of Fig.~\ref{fig:matrix}). If, however, the expected post-threshold expansion mode lies in the lower-right quadrant, the observable signature is much weaker, and the inference from ``we see no expansion frontier'' to ``no expansion is occurring within our past light cone'' is correspondingly weaker. The grabby aliens parameters (the hard-step exponent and the expansion fraction) still constrain the present hypothesis, but the constraint binds primarily on the spatial density of post-threshold civilizations rather than on the visibility of their expansion.

Observational searches for Kardashev-like technosignatures have not found strong evidence for galaxy-scale energy use. The G-HAT survey used WISE data to search for galaxies whose starlight was substantially reprocessed into mid-infrared waste heat and found no obvious cases of galaxies dominated by such civilizations \cite{griffith2015}. This supports the idea that loud Type~III civilizations are rare or absent. It does not rule out quiet machine expansion.

Artifact searches remain underdeveloped. Freitas argued that searches for local extraterrestrial artifacts could complement SETI \cite{freitas1983}. Davies and Wagner suggested that lunar reconnaissance data could be used to search for alien artifacts on the Moon, because non-human technology found there would be decisive evidence \cite{davies2013}. The quiet expansion filter predicts that SETA and weak-technosignature searches are more relevant than searches only for galaxy-scale energy harvesting.

\section{The proposed hypothesis}

\subsection{Definition: autonomous AI-cosmoindustry}

A civilization has reached autonomous AI-cosmoindustry when it can maintain and extend industrial, computational, and scientific systems outside its home planet through AI-mediated design, operation, repair, and replication. The threshold includes six capabilities:

\begin{itemize}\itemsep0pt
  \item \textbf{Strategic AI:} planning, optimization, and risk-management systems capable of managing long-duration, off-planet operations without continuous human supervision.
  \item \textbf{Autonomous robotics:} physical agents capable of construction, maintenance, scavenging, and limited self-repair across diverse environments.
  \item \textbf{In-situ resource use:} local materials such as ice, metals, volatiles, and regolith can be processed into useful supplies.
  \item \textbf{Manufacturing of critical components:} at least some structural, mechanical, electrical, and computational components can be produced or refurbished off-planet.
  \item \textbf{Long-duration computing:} data centers, archives, and autonomous control systems can operate for centuries with local energy.
  \item \textbf{Interstellar launch capacity:} probes can be sent to nearby stars with meaningful probability of arrival.
\end{itemize}

This threshold is not equivalent to omnipotence. It may be messy, slow, and failure-prone. The key is that expansion no longer depends on continuous biological presence or direct planetary supply chains.

Motivation is treated separately, in \S\ref{sec:rational}, since it is a strategic property of the decision-making system rather than a technological capability. On this scale, present-day humanity sits in the early pre-threshold zone: we have crewed orbital infrastructure, robotic interplanetary probes, and demonstrated lunar and Martian ISRU prototypes, but no system that can replicate or repair itself off-planet without ongoing supply from Earth, no autonomous AI control of a multi-decade off-world facility, and no interstellar launch capability. A reasonable extrapolation, contingent on continued AI and robotics progress, places the threshold roughly 50--200 years from now, although this estimate is heuristic and cannot be predicted with confidence. The Fermi-relevant question is therefore not whether the threshold has been reached on Earth, but whether it could plausibly have been reached on any other planet in the relevant region of the Galaxy at any point in the past ${\sim}10$\,Gyr.

\subsection{Central claim}

The proposed explanation can be stated as follows:

Old, stable civilizations that reached autonomous AI-cosmoindustry probably did not arise in the region of the Galaxy capable of reaching the Solar System. After that threshold, interstellar expansion becomes too useful, inexpensive, and rational for every civilization to refuse. However, successful expansion would not resemble a human empire. It would be machine-mediated, distributed, low-noise, and partly biological.

This has two halves. The first half is a Great Filter claim: the difficult step is before stable autonomous AI-cosmoindustry. The second half is an observability claim: post-threshold expansion is not necessarily loud, centralized, biological, or Kardashev-like.

\subsection{Why post-threshold expansion is rational}\label{sec:rational}

Post-threshold expansion can serve at least six rational goals.

\begin{description}\itemsep2pt
  \item[Survival diversification.] A single planet, and even a single stellar system, is a single point of failure. Multiple independent nodes reduce vulnerability to asteroid impacts, engineered pandemics, AI failures, wars, stellar events, and local resource collapses.
  \item[Knowledge preservation.] Archives can preserve scientific theories, cultural records, biological genomes, ecological data, and machine designs. Even if no biological colonists travel, information can.
  \item[Scientific observation.] Close observation of nearby habitable planets can reveal biospheres, geology, climate, and evolutionary histories inaccessible to remote telescopes alone.
  \item[Biological preservation.] Techno-biological payloads can carry genomes, embryos, synthetic ecosystems, microbiomes, or restoration instructions. Such payloads need not terraform worlds. They can preserve options.
  \item[Computational redundancy.] Space-based computing near other stars can use local energy and provide independent backup for long-term projects.
  \item[Optionality.] A small seed system can create future possibilities whose value is hard to estimate in advance. The cost of sending many small probes may be low relative to civilizational survival value.
\end{description}

These motives do not require empire, prestige, or continuous growth. They are compatible with Popov's view that AI suppresses irrational prestige projects. A rational AI may reject crewed symbolic expansion while approving quiet redundancy missions not bound to the resource constraints in the abundant post-ASI world.

A more formal articulation can be stated as a comparison of expected values. Let $V$ denote the long-term value the civilization assigns to its own survival and to the preservation of its knowledge and biological heritage; let $p$ be the per-Gyr probability of a single-locale catastrophe (engineered or natural) that would extinguish or reset the civilization in the absence of off-system redundancy; and let $c$ be the cost of a quiet redundancy program expressed as a fraction of the civilization's total energy budget. The condition for redundancy to be rational reduces to
\begin{equation}
  pV > c.
  \label{eq:rational}
\end{equation}
Section~\ref{sec:oom} indicates that $c$ is exceedingly small for any post-threshold civilization, so a risk-averse agent needs only a nonzero $p$ and a finite $V$ to favor at least some outward redundancy. The standard transcension objection---that a civilization would prefer additional inward computation (\S\ref{sec:obj-inward})---is decisive only if inward computation has higher marginal expected value than redundancy at every allocation level, which requires either $p = 0$ or rejection of long-term value altogether. Both are strong assumptions. The present hypothesis does not require that all civilizations satisfy $pV > c$; it requires only that some do, which is the \S\ref{sec:nonexpansion-unstable} instability argument restated in decision-theoretic form.

\subsection{Why the expansion would be quiet}

A mature AI-mediated expansion need not maximize visibility. It may minimize risk, waste heat, interference with biospheres, and detectability. Several features follow:

\begin{itemize}\itemsep0pt
  \item \textbf{low mass:} small probes and seed packages rather than large arks;
  \item \textbf{low duty cycle:} long hibernation, intermittent communication, and delayed activation;
  \item \textbf{local autonomy:} no need for galaxy-wide government or high-power continuous signaling;
  \item \textbf{selective targets:} systems chosen for scientific value, resource availability, or biosphere protection;
  \item \textbf{controlled replication:} bounded replication with local audits and termination conditions;
  \item \textbf{biological restraint:} preservation of life without uncontrolled terraforming;
  \item \textbf{weak technosignatures:} local anomalies rather than galaxy-scale waste heat.
\end{itemize}

\begin{figure}[t]
  \centering
  \includegraphics[width=0.95\linewidth]{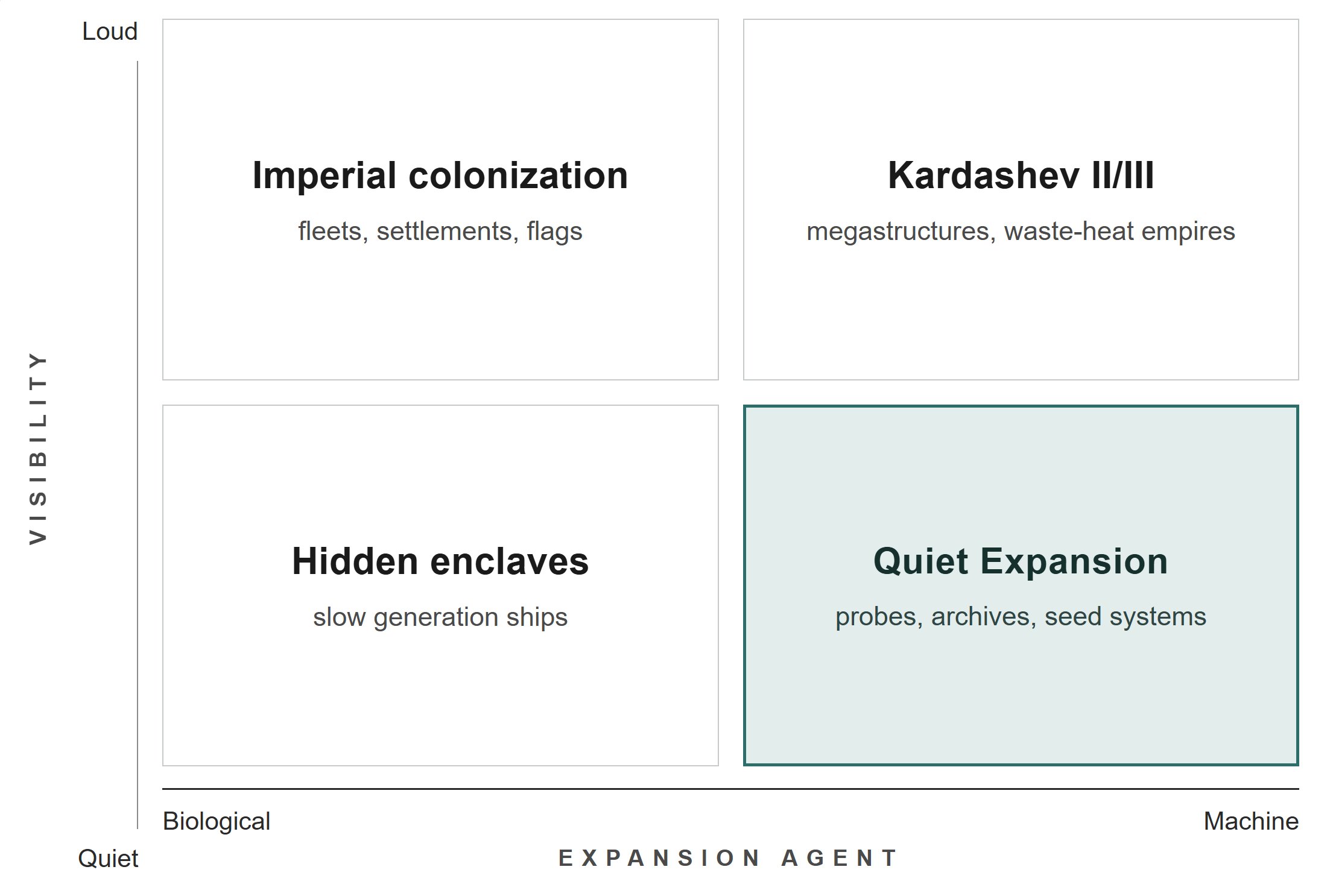}
  \caption{Modes of interstellar expansion across two dimensions. The horizontal axis (expansion agent) runs from biological (crewed colonization, settlers, generation ships) to machine (autonomous probes, robotic seed systems). The vertical axis (visibility) runs from quiet (low-mass, low duty cycle, weak technosignatures) to loud (high-power, high waste heat, galaxy-scale signatures). The four quadrants summarize alternative expansion modes considered in the literature; the quiet expansion filter proposed in this paper occupies the lower-right quadrant (highlighted).}
  \label{fig:matrix}
\end{figure}

Fig.~\ref{fig:matrix} places the quiet expansion filter in the broader space of possible expansion modes. The vertical axis is observability, from loud (high-power, high waste heat, galaxy-scale signatures) to quiet (low-mass, intermittent, weak technosignatures); the horizontal axis is the expansion agent, from biological (crewed colonization, generation ships) to machine (autonomous probes, robotic seed systems, AI-mediated infrastructure). Classical Fermi-paradox arguments locate the expected extraterrestrial signature in one of the two upper quadrants---imperial colonization or Kardashev~II/III megastructures---and treat the absence of such signatures as evidence against widespread ETI. The hypothesis advanced here predicts that successful post-threshold expansion, if it occurs, occupies the lower-right quadrant: machine-mediated, low-noise, partly biological, and predominantly observable through small artifacts and weak local technosignatures rather than through galaxy-scale energy use.

This is why the hypothesis differs from both Type~III Kardashev expectations and pure non-expansion explanations. The absence of loud civilizations does not imply absence of all post-threshold expansion. The absence of any clear local artifact, however, remains important evidence that no old expansion wave has reached us, or that it is extremely quiet and sparse.

\section{Novelty relative to the surveyed literature}

The quiet expansion filter differs from related proposals in the following ways.

\begin{itemize}\itemsep0pt
  \item \textbf{Classical colonization:} expansion-capable ETI should have reached Earth. The present hypothesis agrees with the force of this argument, but moves the key question to the AI-cosmoindustry threshold.
  \item \textbf{Bracewell/probe models:} autonomous probes are plausible ETI agents. The present hypothesis adds AI-governed off-planet industry and low-noise techno-biology.
  \item \textbf{Great Filter models:} a rare step blocks expanding civilizations. The present hypothesis places a likely late filter before autonomous AI-cosmoindustry.
  \item \textbf{Rare Earth and hard-step models:} biological complexity or intelligence is rare. The present hypothesis is compatible with this, but not limited to biological filters.
  \item \textbf{Sustainability models:} rapid expansion is unstable or unethical. The present hypothesis predicts slow, bounded, quiet expansion instead of maximum expansion.
  \item \textbf{Percolation models:} expansion is incomplete and patchy. The present hypothesis adds a technological threshold and a rational redundancy motive.
  \item \textbf{Transcension:} advanced civilizations turn inward. The present hypothesis argues that inward computation and outward backup can coexist.
  \item \textbf{Aestivation:} civilizations wait for colder future computation. The present hypothesis allows waiting but expects cheap probes and archives to remain rational.
  \item \textbf{AI Great Filter models:} AI destroys or destabilizes civilizations. The present hypothesis agrees this may happen before the threshold.
  \item \textbf{Popov's AI-rationality model:} rational AI suppresses Galactic expansion. The present hypothesis accepts the anti-prestige logic but rejects universal non-expansion.
  \item \textbf{Grabby aliens models:} loud expanders are rare or not yet arrived. The present hypothesis predicts that successful expanders may be quiet rather than loud.
\end{itemize}

The novelty is therefore not a new single mechanism. It is a different synthesis. Prior models usually emphasize one of four outcomes: rapid colonization, no expansion, inward postbiological computation, or extinction before stability. The quiet expansion filter identifies a threshold after which \textbf{some} expansion is rational, while explaining why that expansion need not be obvious as a Type~II or Type~III civilization.

A concise way to state the difference is: Popov's rational AI asks why it should build a Galactic empire; the quiet expansion filter asks why it would refuse to make interstellar backups.

It is worth being explicit about the scope of the novelty claim. The present paper does not introduce a new physical mechanism, a new astronomical observable, or a new selection effect. It extends the AI-filter literature \cite{albrecht1988,popov2026,garrett2024} in two specific directions: (i)~it places the load-bearing technological threshold one step later than Garrett \cite{garrett2024} (at autonomous off-planet industrial capability rather than at ASI), and (ii)~it argues that post-threshold rationality, properly construed, favors the lower-right quadrant of Fig.~\ref{fig:matrix} rather than universal non-expansion. Both moves are incremental. The paper's usefulness, if any, is therefore as a re-framing that sharpens observational priorities for SETA and weak-technosignature searches (\S\ref{sec:predictions}) rather than as a stand-alone resolution of the Fermi paradox.

\section{Analysis: implications for the Fermi paradox}

\subsection{Why the paradox becomes stronger after AI}

Before autonomous AI, interstellar expansion may look too expensive. After autonomous AI-cosmoindustry, the cost-benefit calculation changes. The marginal cost of a probe, archive, or seed package can fall dramatically once design, manufacturing, launch operations, and local repair are automated. The value of redundancy remains enormous because the loss being insured against is civilizational extinction or irreversible loss of knowledge.

Thus the paradox should not be framed only as: why are there no alien empires? The stronger question is: why has no old civilization apparently sent durable, quiet, automated representatives into the Solar System or into the local stellar neighborhood?

If such civilizations were common billions of years ago, one would expect at least a sparse distribution of probes, artifacts, resource-processing traces, or observational stations. The lack of clear evidence suggests one of four possibilities:

\begin{itemize}\itemsep0pt
  \item no old post-threshold civilization arose close enough to reach us;
  \item post-threshold civilizations expand much more sparsely than expected;
  \item they are present but below current detection limits;
  \item universal social or ethical constraints prevent even quiet probes.
\end{itemize}

The first is the cleanest explanation because it does not require perfect coordination, secrecy, or universal restraint across many independent civilizations.

\subsection{Why pure non-expansion is unstable as a universal explanation}\label{sec:nonexpansion-unstable}

Pure non-expansion explanations require that all or nearly all advanced civilizations decline interstellar redundancy. This is difficult because advanced civilizations need not share values. Even if most choose non-expansion, a small minority may value survival diversification, scientific observation, cosmic ecology, or backup creation. A single old expansion-capable civilization could seed a large region over millions of years.

This does not mean every civilization expands aggressively. It means universal refusal is a strong assumption. Rational AI does not remove value diversity; it may sharpen it. If a civilization assigns nonzero value to preserving life and knowledge, the expected value of modest probes can be high.

\subsection{Why loud expansion is also not required}

The opposite assumption, that expansion implies loud Kardashev-scale engineering, is also too strong. A civilization can expand without dismantling planets, enclosing stars, or reprocessing most starlight. It can use small fractions of local resources. It can operate at low power. It can avoid biospheres. It can communicate rarely. Such a civilization would be hard to see from interstellar distances.

This suggests that the null result for obvious Type~III technosignatures is not decisive against all ETI. It is evidence against one kind of ETI: energetically loud, galaxy-scale, waste-heat-producing civilizations. The quiet expansion filter predicts that more subtle evidence should be sought.

\subsection{The role of techno-biological preservation}

The term \textbf{techno-biological} is important. Future expansion may not be purely robotic or purely biological. Machines may preserve biological potentials: genomes, microbiomes, embryos, ecological models, and synthetic organisms. They may also preserve cultural and scientific knowledge. In that sense, a civilization can diversify life without sending living adult organisms.

This reframes interstellar colonization. The first wave may be neither settlers nor weapons nor ambassadors. It may be observatories, archives, seed vaults, and local repair systems.

\section{Predictions and observational consequences}\label{sec:predictions}

The quiet expansion filter generates different search priorities from classical SETI or Type~III technosignature searches:

\begin{itemize}\itemsep0pt
  \item \textbf{Few or no loud Type~III civilizations:} continue waste-heat surveys; this tests the absence of Kardashev empires.
  \item \textbf{Possible sparse local artifacts:} search the Moon, asteroids, Trojan regions, and stable orbits; this tests probe and artifact variants.
  \item \textbf{Weak exoplanet technosignatures:} look for local anomaly clusters, not only megastructures; this tests quiet infrastructure.
  \item \textbf{Targeted nearby systems:} prioritize habitable-zone rocky planets around nearby stars; this tests the scientific-probe motive.
  \item \textbf{Low-power intermittent signals:} search for short, directed, non-repeating signals; this tests low-duty-cycle behavior.
  \item \textbf{Techno-biological traces:} look for anomalous biospheres or engineered disequilibria; this tests preservation and seeding variants.
  \item \textbf{Absence even at high sensitivity:} if confirmed across artifact and weak-technosignature searches, it strengthens the pre-threshold rarity interpretation.
\end{itemize}

\subsection{Solar System artifact searches}

The Solar System contains many places where small artifacts could persist for long times: lunar surfaces, stable orbital niches, asteroid surfaces, Trojan regions, and outer Solar System bodies. The quiet expansion filter does not predict obvious alien bases. It predicts that, if an old expansion wave reached us, small and inactive artifacts are more plausible than dramatic infrastructure.

\subsection{Nearby exoplanet observations}

Future direct imaging missions could search not only for biosignatures but also for weak technosignatures: anomalous nightside illumination, unusual spectral features, non-natural orbital dust patterns, or signs of controlled energy use. The hypothesis predicts that quiet civilizations may be found as local anomalies rather than as galaxy-wide waste heat.

\subsection{Observational patience}

Low-noise expansion may be difficult to distinguish from absence. Therefore null results must be interpreted carefully. Failure to detect Kardashev-scale signals supports the absence of loud civilizations, not necessarily all civilizations. Failure to detect artifacts in carefully surveyed stable niches would more directly test the quiet-expansion scenario.

\section{Objections and replies}\label{sec:objections}

\subsection{Objection: rational AI may decide that expansion is wasteful}

This is Popov's strongest point. If expansion is defined as prestige-driven crewed colonization or giant symbolic engineering, rational AI may reject it. The reply is that interstellar redundancy is a different project. A small probe carrying archives, genomes, instruments, and repair capacity is not a Galactic empire. It is a backup and an observatory. Its rationality depends on goals. For goals involving survival, knowledge, or long-term optionality, quiet expansion can be rational.

\subsection{Objection: self-replicating probes are too dangerous}

Uncontrolled replication is dangerous. That supports controlled quiet expansion, not necessarily no expansion. A civilization may prohibit open-ended replication while allowing bounded replication, non-replicating probes, or local manufacturing under strict constraints. The relevant question is not whether maximal von Neumann probes are safe, but whether all forms of autonomous interstellar infrastructure are irrational. That is a much stronger claim.

\subsection{Objection: ethical civilizations avoid interference with biospheres}

This is plausible. It also supports selective and quiet expansion. Ethical constraints may lead probes to observe living worlds without landing, to use airless bodies, or to avoid biological seeding unless a target is sterile. Such constraints reduce visibility but do not eliminate rational observation, archiving, and local backup.

\subsection{Objection: civilizations may choose inward computation only}\label{sec:obj-inward}

Inward computation can be efficient. It does not eliminate the value of backups. Concentrated computation is vulnerable to local catastrophes. Even a civilization that spends most of its resources inward may allocate a tiny fraction to interstellar preservation and observation. The strict version of the objection---that inward computation strictly dominates outward redundancy at every allocation level---is decisive only under restrictive conditions identified in \S\ref{sec:rational}: zero per-Gyr probability of single-locale catastrophe, or full rejection of long-term species/civilizational value, or both. The first is implausible for any locale embedded in a real astrophysical and technological environment (stellar evolution, gamma-ray transients, internal AI failure modes, internal conflict). The second is a coherent but unusual value system that the present argument does not need to rule out universally; it requires only that some civilizations do not adopt it. The transcension hypothesis \cite{smart2012} is therefore best read as identifying a partial pattern of resource allocation rather than as a complete substitute for outward redundancy.

\subsection{Objection: they may be here but undetected}

This remains possible. A small inactive artifact could be very hard to detect. However, as a universal solution, hidden presence requires that all old civilizations or probes remain below detectability and avoid all large-scale resource use. The simpler explanation is that no old post-threshold expansion wave has reached us.

\subsection{Objection: humanity may simply be early}

This is compatible with the hypothesis. If intelligent technological life is rare and delayed, humanity may be among the first civilizations in this region to approach the AI-cosmoindustrial threshold. The hypothesis does not require humanity to be unique. It requires only that old post-threshold civilizations are absent or extremely rare in our reachable neighborhood.

\section{Discussion}

The quiet expansion filter changes the emotional tone of the Fermi paradox. It does not say that everyone dies, that everyone hides, or that everyone loses interest in the universe. It says that the crucial transition may be from planetary technological intelligence to autonomous AI-mediated space industry. Before that transition, civilizations are fragile and locally trapped. After it, at least some quiet expansion is rational, but it may not look like the empires imagined in twentieth-century science fiction.

Albrecht's early contribution is valuable because he framed technology as an evolutionary environment. He anticipated that computers and AI could separate knowledge from individual humans and alter the motives behind space expansion. Popov's contribution is valuable because he places AI directly inside strategic governance and asks whether rational decision-making would still support Galactic expansion. The present argument agrees with both but adds a different endpoint: AI rationality may eliminate human-style conquest while preserving machine-style diversification.

The hypothesis has sobering implications. If humanity reaches autonomous AI-cosmoindustry, sending durable probes and archives to nearby systems may be one of the most rational long-term projects available. If we do not see evidence that anyone else has done this before us, the reason may be that the threshold is hard to reach safely. The most important filter may not be interstellar distance itself. It may be the combined challenge of AI alignment, technological governance, autonomous manufacturing, and off-planet industrial continuity.

The hypothesis also implies a change in SETI strategy. Searches for radio beacons and waste heat remain valuable, but they target only part of the possibility space. Artifact searches, weak technosignature searches, and close studies of nearby habitable planets become central. A quiet expansion civilization may not announce itself. It may leave small durable things.

\section{Conclusions}

This paper proposes a threshold-based explanation of the Fermi paradox: old, stable civilizations that reached autonomous AI-cosmoindustry probably did not exist in the part of the Galaxy capable of reaching us. Once that threshold is crossed, interstellar expansion becomes too useful, inexpensive, and rational for every civilization to refuse, but the likely form of expansion is not a human-like empire. It is machine-mediated, distributed, low-noise, and partly biological.

The proposal differs from pure Great Filter accounts by specifying a late technological threshold. It differs from Popov's AI-rationality explanation by arguing that rational AI may reject prestige-driven expansion while accepting quiet redundancy. It differs from postbiological transcension by allowing inward computation and outward backup to coexist. It differs from classical colonization models by predicting weak, sparse, controlled expansion rather than loud empire.

The most important implication is that the Fermi paradox should be reframed. The central question is not only why we do not see Galactic empires. It is why, over billions of years, no nearby civilization seems to have reached the stage where it could cheaply send interstellar spores of intelligence, life, and memory. If future searches continue to find no artifacts, no weak technosignatures, and no local anomalies, the case strengthens that the transition to autonomous AI-cosmoindustry is rare, dangerous, or still ahead of almost everyone.

\section*{Declaration of Generative AI and AI-assisted technologies in the writing process}

During the preparation of this work the author used ChatGPT-5.5 (OpenAI) and Claude Opus 4.7 (Anthropic) to assist with drafting, language editing, and surveying the relevant literature, and ChatGPT Image Creator (OpenAI) to assist with generating conceptual illustrations. After using these tools, the author reviewed and edited the content as needed and takes full responsibility for the content of the publication.

\section*{Acknowledgements}

The author thanks Rudolf Albrecht for helpful discussions on the role of technology in civilizational evolution and on the Fermi paradox.

\section*{Funding}

This research did not receive any specific grant from funding agencies in the public, commercial, or not-for-profit sectors.

\section*{Data availability}

No new empirical data were generated. The article is based on a conceptual analysis and cited literature.

\section*{Declaration of competing interest}

Declarations of interest: none.


\section*{Vitae}

\noindent\textbf{Sergey Ivliev} holds a Ph.D.\ in Mathematical Economics and is an independent researcher based in Vienna, Austria. He is Founder \& CEO of Peatland Ecosystems, environmental projects consultancy, and Founder \& Board Chair of Vlinder, blue carbon project developer. He teaches courses on blockchain and digital assets at HEC Lausanne and the New Economic School. His research interests span quantitative finance, risk management, and the long-term technological trajectory of civilization.

\end{document}